\documentstyle[12pt]{article}
\begin{document}

\begin{flushright}
Preprint IP-ASTP-19; FIAN-TD 94/7\\
August 1994
\end{flushright}
\vspace{13mm}
\begin{center}
{\Large\bf CUMULANTS OF QCD \\ 
\vspace{1mm}
MULTIPLICITY DISTRIBUTIONS \\ 
\vspace{3mm}
IN SMALL PHASE SPACE BINS}\\

\vspace{8mm}
{{\large\bf I.M. DREMIN}\footnote{Permanent address: Lebedev Physical Institute, Leninsky prospect, 53, Moscow 117924, Russia. E-mail-address: dremin@td.lpi.ac.ru } \\
\vspace{3mm}
Institute of Physics, Academia Sinica \\
 Taipei, Taiwan 11529, The Republic of China} 
\end{center}


\vspace{5mm}
\begin{abstract}
It is shown that, as functions of their rank, cumulants of QCD multiplicity distributions in small phase space bins possess the quasi-oscillating behavior 
similar to that found for them in the total rapidity range. First minimum moves to lower ranks for smaller bins. For the total rapidity range, it moves to higher ranks with energy increase. The running property of the QCD coupling constant is in charge of these effects which can be verified in experiment.
\end{abstract}

Recently, it was shown [1] (for a review, see [2]) that cumulants of multiplicity distributions in QCD behave in a rather peculiar way revealing pronounced oscillations. The analysis was done for the total rapidity range.
Its predictions have been confirmed by experiment [3]. From the theoretical side, those results imply that QCD distributions do not belong to the class of infinitely-divisible ones (in particular, they can not be fitted by the negative binomial distribution - NBD) and prohibit Poissonian cluster models [4].

Here, we shall answer two more questions related to the behavior of cumulants. First, we show that the first minimum of the ratio of cumulant to factorial moments for the total rapidity range moves to higher ranks when the primary energy increases. Second, we show that the similar minimum appears when multiplicity distributions in small phase space bins are analyzed\footnote{Such an analysis has become very popular in connection with ideas of intermittency and fractality (for a review, see [5]).} and it shifts to lower ranks for ever smaller bins. 

The above properties can be easily guessed from the formulae obtained in the original paper [1]. Therefore, by using them, we shall, first, give qualitative arguments in favor of such properties, then show how different factors influence the conclusion, and, finally, discuss its experimental implications.

In its simplest double-logarithmic approximation (DLA), quantum chromodynamics fails to describe experimentally observed multiplicity distributions predicting very wide shape of them. However, already in the next (modified leading logarithm - MLLA) approximation its predictions become quite reasonable.
Besides considering just the multiplicity distributions $P_n$, it is convenient to use their factorial ($F_q$) and cumulant ($K_q$) moments defined by the relations
\begin{equation}
F_q = \frac {\langle n(n-1)...(n-q+1)\rangle }{\langle n\rangle ^{q}} = \frac {\sum n(n-1)...(n-q+1)P_n }{(\sum nP_n )^q } ,
\end{equation}
\begin{equation}
F_q = \sum _{m=0}^{q-1} C_{m}^{q-1}K_{q-m}F_{m} ,
\end{equation}
where $C_{m}^{q-1}$ are the binomial coefficients, and the set of recurrence relations (2) is applied to define the cumulants when the factorial moments are determined according to eq.(1).

As a function of the rank $q$, their ratio
\begin{equation}
H_q = \frac {K_q}{F_q}
\end{equation}
behaves like $q^{-2}$ in DLA\footnote{To get an insight, let us notice that the negative binomial distribution with a parameter $k$ interpreted as a number of independent sources predicts the ratio $H_{q}^{NBD} = q^{-k}$. Herefrom, one concludes that DLA corresponds to NBD with rather small value of the parameter $k = 2$ and, consequently, gives rise to very wide multiplicity distribution.} but acquires a minimum in MLLA crossing once the abscissa axis. Higher-order approximations predict further crossings and quasi-oscillatory behavior of the ratio (see [2]). The position of the first minimum, already observed in MLLA, seems rather stable. It depends on the QCD coupling constant only and is not strongly influenced by the higher-order terms. In gluodynamics, it is given [1] approximately as
\begin{equation}
q_{min}\approx \frac {1}{h_{1}\gamma _{0}} + \frac {1}{2} + O(\gamma _{0}) ,
\end{equation}
where $\gamma _{0} = (6\alpha _{S}/\pi )^{1/2} , \alpha _{S}$ is the QCD coupling constant, and $h_{1} = 11/24$ is the coefficient calculable in MLLA. Higher approximations contribute terms $O(\gamma _{0})$, and, in practice, can be neglected when dealing with the position of the very first minimum.

The formula (4) immediately shows to us that the position of the first minimum moves to higher values for jets with larger virtuality $Q$ since the QCD coupling constant is running as $\ln ^{-1}Q$. Therefore, the predicted shift of the minimum is 
\begin{equation} 
q_{min} \propto \ln ^{1/2}Q
\end{equation}
with increase of virtuality $Q$. Thus, analyzing the multiplicity distributions at ever higher energies one would notice the shift of the minimum of $H_q$ to higher rank values.

One could guess already that the position of the minimum would shift to smaller values of $q_{min}$ if multiplicity distributions within small phase space bins are considered. Really, it would mean considering the subjets with lower virtuality, and, therefore, with larger effective values of $\alpha _{S}$ and $\gamma _{0}$. To confirm this statement and estimate the various factors, contributing to it, one should perform the analysis anagolous to that done in [6] where factorial moments for small phase space bins were treated. Referring for further details to [6], let us outline the main steps of the analysis.

To calculate the moments of the distributions within small bins, one should convolute three stages of the process of the emission of a gluon jet by a quark:
\\
\begin{enumerate}
\item the initial emission of a hard primary gluon jet by a quark, 
\item its subsequent evolution giving rise to secondary jets with some energy spectrum, 
\item the choice (among those jets) of the final subjet which hits the bin under consideration and the estimate of its correlators. 
\end{enumerate}
It has been done in [6] for factorial moments of the distributions in small phase space bins but can be repeated with slight modifications for cumulants as well. For example, for cumulants one writes down the formula analogous to (3.2) of [6] :
\begin{eqnarray}
K_{q}\left( Q\Theta _{0}; \frac {\Theta _{0}}{\Theta }\right) \langle n(Q\Theta _{0})\rangle ^{q}&\propto &\int ^{Q}\frac {dE}{E}\frac {\alpha _{S}}{2\pi }\Phi _{F}^{G}\left( \frac {E}{Q}\right) \int ^{E}\frac {dk}{k}D^{\Theta }\left( \frac {E}{k};E\Theta _{0};k\Theta \right) \nonumber \\&        &\langle n(k\Theta ;1)\rangle ^{q}H_{q}(k\Theta ;1) ,
\end{eqnarray}
where the energy $E$ integration with Altarelli-Parisi kernel $\Phi $ corresponds to the stage 1), the energy $k$ integration with the spectrum $D^{\Theta }$ is according to the stage 2), and the factor $\langle n\rangle ^{q} H_q$ describes the cumulant of the subjet hitting the angular window $\Theta $ (the stage 3)). Here $Q = E_{0}\Theta _{0},  E_0 $ is the primary jet energy and $\Theta _{0}\sim 1$. The only difference from the integrals for factorial moments in [6] is due to the factor $H_q$ for the final subjet which should be evaluated. However, this factor is a slowly varying function of $k$ while the spectrum $D^{\Theta }$ and the $q$-th power of multiplicity change very rapidly. Therefore, the steepest-descent method can be again applied to calculate the correlator (6) with $H_q$ estimated at the stationary point. Since the unnormalized factorial moments are given by the similar expressions without $H_q$ in the integrals, one gets, after dividing cumulants by factorial moments, that the ratio $H_q$ for the multiplicity distribution in a small bin is just the ratio for the subjet hitting the bin when calculated at the stationary point $\kappa _{0} = (k\Theta )_{s.p.}$. Because the subjet products cover completely the bin $\Theta $, its ratio $H_q$ is to be calculated from the same equations as for the total rapidity range\footnote{The only difference is that the total phase space should be replaced by the phase space window under consideration.} and, therefore, possesses the similar quasi-oscillatory behavior. However, the positions of maxima and minima are shifted because it is calculated now at the stationary point $\kappa _{0}$ but not for the total range of angles.

We shall deal with the behavior of the first minimum which is well described by MLLA where the ratio $H_q$ is given [1] by the equation
\begin{equation}
H_{q}^{MLLA} = \frac {\gamma _{0}^{2}(1-2h_{1}q\gamma )}{q^{2}\gamma ^{2}} .
\end{equation}
Here, $\gamma $ is an anomalous dimension of QCD determining the energy dependence of mean multiplicity by the relation $\langle n\rangle = \exp (\int  dt\gamma (t))$.
The only problem now is to insert $\gamma $ for small bins and to use the stationary point value of $\gamma _{0}$. Actually, the value of $\gamma $ in MLLA for small bins was calculated in [6] (see eq.(3.30) there) and, in case of gluodynamics, is given by
\begin{equation}
\gamma = \gamma _{0}(1-c_{q}\gamma _{0}) ,
\end{equation}
\begin{equation}
c_q = \frac {11}{48}\frac {q^3 - 3q^2 - q - 1}{(q^2 + 1)(q + 1)} .
\end{equation}
One easily gets that in the region of interest to us $q = 3-5$ the correction term is negligible
\begin{equation}
c_{q}\gamma _{0}\ll 1 .
\end{equation}
Therefore, for our purposes $\gamma \approx \gamma _{0}$, and one gets the result (4) with high enough precision. Thus, MLLA modification of the anomalous dimension $\gamma $ is not important here.

The stationary point of the integrand in eq.(6) is estimated using the standard methods [6, 7]. For qualitative estimates, we are aimed to, it is enough to demonstrate the behavior of $\gamma _{0}$ at the stationary point as calculated for fixed coupling since influence of its running on the position of the stationary point is a higher-order effect. Then it is easy to get for the virtuality $\kappa _{0}$ of the subjet at that point
\begin{equation}
\kappa _{0} = Q \left( \frac {\Theta }{\Theta _{0}}\right) ^{1+q^{-2}} ,
\end{equation}
i.e. "the effective mass" of the subjet diminishes almost proportionally to the opening angle $\Theta $ with additional (and not important for our purposes) factor $\Theta ^{1/q^{2}}$ appearing due to the energy spectrum $D^{\Theta }$ decrease. Therefore, the slight shift of the minimum position $q_{min}$ to smaller values is envisaged at small bins since $q_{min}\propto \gamma _{0}^{-1}$ contains the term $\sim \ln ^{1/2}\kappa _{0}$ (let us note that $\kappa _{0}$ replaces $Q$ compared to eq.(5)).

Even though more rigorous expressions are available, we decided to use the semiquantitative description due to two reasons. First, mathematical details are rather cumbersome but do not change main conclusions presented above. Second, the experimental verification of the results is also possible at the semiquantitative level only so that the theoretical precision should not exceed the experimental facilities. That statement is easy to appreciate if one reminds that experimentalists can get the ratio $H_q$ at integer values of $q$ only. They proceed from the factorial moments evaluated according to eq.(1) to the cumulants defined through the set of recurrence relations (2) for integer points $q$ . Therefore, they can not observe the continuous drift of $q_{min}$ with energy or with bin sizes. Nevertheless, such a shift (if noticed as a shift of the minimum position from one integer rank value to another) will be of interest since it demonstrates further details predicted by QCD.

 Probably, the best way to find it out in experimental data is to study near the minimum point $q_{min}$ the evolution of the differences $F_{q}-F_{q-1}$ with increase of energy and/or with decrease of bin sizes. Unfortunately, it is hard to measure high-rank cumulants at small bins with high enough precision. Besides, they are very sensitive to the selection procedure in a given experiment (even though this sensitivity can be, probably, used to compare the criteria attempted by different experimental groups).

In conclusion, we have shown that the minimum of the ratio of cumulant to factorial moments of multiplicity distributions predicted by QCD [1] and confirmed by experiment [3] is subject to the shift to higher values of ranks at higher energies and to smaller ranks when multiplicity distributions in small phase space bins are measured. Experimental verification of those effects would demonstrate once again the running property of the QCD coupling, and, furthermore,  the very applicability of QCD to description of subtle correlation effects even for rather soft processes of multiparticle production.

\vspace{1mm}
\begin{center}
\large{Acknowledgements}
\end{center}
\vspace{1mm}

I am grateful to S.-C. Lee for the invitation to visit the Institute of Physics 
of Academia Sinica where this work was done. \\
This work is supported by the Russian Fund for Fundamental Science (grant 94-02-3815) and by the International Science Foundation.

\newpage
\begin{center}
\large{\bf REFERENCES}
\end{center}
\vspace{2mm}

[1] I.M. Dremin, Phys.Lett.B 313 (1993) 209.\\

[2] I.M. Dremin, Physics-Uspekhi 164(8) (1994) 3. \\

[3] I.M. Dremin, V. Arena, G. Boca et al, Phys.Lett.B (1994). \\

[4] S. Hegyi, Phys.Lett.B 327 (1994) 171.\\

[5] E.A. DeWolf, I.M. Dremin and W. Kittel, Usp.Fiz.Nauk 163(1) (1993) 3;
Phys.Rep. (to be published).\\

[6] Yu.L. Dokshitzer and I.M. Dremin, Nucl.Phys.B 402 (1993) 139.\\

[7] Yu.L. Dokshitzer, G. Marchesini and G. Oriani, Nucl.Phys.B 387 (1992) 675.

\end{document}